\documentstyle[12pt]{article}
\def\beq{\begin{equation}}
\def\eeq{\end{equation}}

\begin{document}

\begin{titlepage}
\begin{center}
{\Large \bf Theoretical Physics Institute \\
University of Minnesota \\}  \end{center}
\vspace{0.3in}
\begin{flushright}
TPI-MINN-92/46-T \\
September 1992
\end{flushright}
\vspace{0.4in}
\begin{center}
{\Large \bf Comment on ``Summing one-loop graphs at multi-particle
threshold"\\}
\vspace{0.2in}
{\bf M.B. Voloshin  \\ }
Theoretical Physics Institute, University of Minnesota \\
Minneapolis, MN 55455 \\
and \\
Institute of Theoretical and Experimental Physics  \\
                         Moscow, 117259 \\
\vspace{0.2in}
{\bf   Abstract  \\ }
\end{center}

The propagator of a virtual $\phi$-field with emission of $n$ on-mass-shell
particles all being exactly at rest is calculated at the tree-level in
$\lambda \phi^4$ theory by directly solving recursion equations for the sum
of Feynman graphs. It is shown that the generating function for these
propagators is equivalent to a Fourier transform of the recently found
Green's function within the background-field technique for summing graphs at
threshold suggested by Lowell Brown. Also the derivation of the result that
the tree-level on-mass-shell scattering amplitudes of the processes $2 \to
n$ are exactly vanishing at threshold for $n > 4$ is thus given in the more
conventional Feynman diagram technique.

\end{titlepage}

The technique recently suggested by Lowell Brown$^{\cite{brown}}$ for
summing the tree graphs in $\lambda \phi^4$ theory for production at the
threshold of $n$ on-mass-shell scalar $\phi$-bosons by a highly virtual
$\phi$ field was extended$^{\cite{v}}$ to calculation of the loop
corrections to the same process. The latter calculation is based on deriving
the two-point Green function of the quantum field $\phi$ in the background
of the complex field of the classical solution to the field equations
considered by Brown. As a by-product of the calculation$^{\cite{v}}$ it was
found that the sum of tree graphs for the processes $2 \to n$ in which all
particles are on the mass shell are exactly vanishing at the threshold for
any $n$ greater than 4 in the $\lambda \phi^4$ theory with unbroken
symmetry. It was subsequently shown$^{\cite{smith}}$ that in the theory with
spontaneous symmetry breaking (i.e. with negative $m^2$) the same behavior
of the tree-level threshold amplitudes holds for any $n$ larger than 2.
Given that the method used in Refs.\cite{brown,v} is not entirely
conventional it is worthwhile to present a derivation of the same results
in more standard terms. It is the purpose of this comment to derive the
Green function equivalent to the one obtained in Ref.\cite{v} as well as the
aforementioned surprising threshold behavior of the on-mass-shell scattering
amplitudes using standard Feynman diagram technique.  Simultaneously this
will substantiate the claim$^{\cite{v}}$ that the technique based on
recursion relations$^{\cite{v1}}$ for the sums of Feynman graphs is
equivalent to Brown's method, though perhaps less elegant.  For simplicity
we concentrate on the case of unbroken $\lambda \phi^4$ theory in which the
Lagrangian reads as

\beq
{\cal L} = {1 \over 2} (\partial \phi)^2 -{1 \over 2} m^2 \phi^2 - {1 \over
4} \lambda \phi^4 ~.
\label{lagrangian}
\eeq
The calculations for the case of the theory with broken symmetry contain no
principal differences from this somewhat simpler in terms of notation case.

Our purpose here is to calculate the propagator $D_n(p)$
of the field $\phi$ with the tree-level
emission of $n$ particles all being at rest and $p$ being the final
four-momentum in the propagator after the emission, see Fig.1. The incoming
four-momentum in the propagator is fixed:

\beq
p_1=p+n q~,
\label{cl}
\eeq
where $q$ is the four-momentum of each of the final particles. In the rest
frame of the produced on-mass-shell particles one has $q_0=m, ~{\bf q}=0$.
For the first two values of $n$ the
propagator is well known:

\beq
D_0(p)={i \over {p^2-1}}~, \hspace{0.8in}
D_2(p)={{i\, 6 \lambda} \over {(p^2-1)((p+2q)^2-1)}}
\label{init}
\eeq
(hereafter we set $m$ equal to one, and it can be also noticed that $n$
is always even due to the unbroken symmetry under the reflection $\phi \to
-\phi$).

The propagators $D_n(p)$ are related by recursion relations analogous to
the ones considered in Ref.\cite{v1}, which graphically are shown on
Fig.2 and algebraically can be written as

\beq
D_n(p)={3 \lambda \over {((p+nq)^2-1)}} \sum_{n_1,n_2,n_3}
\delta_{n_1+n_2+n_3,\,n} {{n!} \over {n_1!\,n_2!\,n_3!}} a(n_1) a(n_2)
D_{n_3}(p)~,
\label{rec1}
\eeq
where $a(N)$ are the amplitudes of threshold production of $N$ particles,
which were found$^{\cite{v1,akp,brown}}$ to be given by:

\beq
a(N)= \langle N | \phi(0) | 0 \rangle = N! \left ( {\lambda \over 8} \right
)^{{N-1} \over 2}
\label{a}
\eeq
where $N$ and thus also $n_1$ and $n_2$ in
eq.(\ref{rec1}) are necessarily odd.  Substituting this expression for the
amplitudes $a(N)$ into eq.(\ref{rec1}) and introducing instead of $D_n(p)$
the normalized propagator $d_n(p)$:

\beq
D_n(p)=i\, n! \left ( {\lambda \over 8} \right )^{n/2} d_n(p)
\label{dn}
\eeq
one can rewrite the equation (\ref{rec1}) in the form

\beq
((p+nq)^2-1) d_n=24 \sum_{n_3{\rm (even)}} {{n-n_3} \over 2} d_{n_3}(p)~.
\label{rec2}
\eeq
This equation can be solved by the technique of generating
function$^{\cite{akp}}$, i.e. by introducing the function

\beq
g_p(x)=\sum_{n=0}^{\infty} x^n d_n(p)~.
\label{gen}
\eeq
Equation (\ref{rec2}) can then be readily verified to be the $n$-th term of
the expansion in $x$ of the differential equation for the generating
function

\beq
\left [ x^2 {{d^2} \over {dx^2}} + (2 (p \cdot q)+1) x {d \over {dx}} +
p^2-1  - {{24 x^2} \over {(1-x^2)^2}} \right ] g_p(x)= 1 ~,
\label{difur}
\eeq
where the inhomogeneous term on the right hand side is fixed by the
normalization of $D_0(p)$. To simplify the subsequent formulas we introduce
the notations: $\epsilon = (p \cdot q)$ and $\omega^2 = (p \cdot q)^2-p^2 +
1$. In the rest frame of the produced on-shell particles $\epsilon = p_0$
and $\omega^2={\bf p}^2+1$. In the kinematics of the process $2 \to n$ i.e.
when the initial and the final momenta in the propagator $D_n(p)$ correspond
to incoming particles the value of $\epsilon$ is negative. The mass shell
for the particle with the momentum $p$ corresponds to $\omega = -\epsilon$
and for the particle with the initial momentum $p+nq$: $\omega =
n+\epsilon$.  Thus $\epsilon=-n/2$ when both incoming particles are on the
mass shell.

Equation (\ref{difur}) is solved by introducing new variable $t$ according
to $x=i e^t$ and then seeking the solution in the form

\beq
g_p(x(t))=e^{-\epsilon t} y_p(t)~.
\label{gy}
\eeq
In terms of the function $y_p(t)$ the equation (\ref{difur}) reads as

\beq
\left [ {{d^2} \over {dt^2}} - \omega^2  + {6 \over {(\rm cosh}\, t)^2}
\right ] y_p(t) = e^{\epsilon t} ~.
\label{difurt}
\eeq
The differential operator in the homogeneous part of this equation is the
same as in eq.(17) of Ref.\cite{v}. Therefore one can use the result for the
Green function from there and write the solution to the equation
(\ref{difur}) in the form

\beq
g_p(x(t))= - { {e^{-\epsilon t}} \over W} \left [ f_1(t)\, \int_{-\infty}^t
e^{\epsilon s}\,f_2(s)\, ds + f_2(t)\, \int_t^\infty e^{\epsilon s} \,
f_1(s) \, ds \right ]~,
\label{solut}
\eeq
where

\beq
f_1(t)={{2 - 3\,\omega + {\omega^2} - 8\,{u^2} + 2\,{\omega^2}\,{u^2}
+ 2\,{u^4} + 3\,\omega\,{u^4} +
{\omega^2}\,{u^4}}\over {{u^\omega}\,{{\left( 1 + {u^2} \right) }^2}}}
\label{f1}
\eeq
and

\beq
f_2(t)={{{u^\omega}\,\left( 2 + 3\,\omega + {\omega^2}
- 8\,{u^2} + 2\,{\omega^2}\,{u^2} + 2\,{u^4} -
 3\,\omega\,{u^4} + {\omega^2}\,{u^4} \right) }\over
{{{\left( 1 + {u^2} \right) }^2}}}
\label{f2}
\eeq
are the two solutions of the homogeneous equation (\ref{difurt}) written in
terms of $u(t)=-i x(t)=e^t$ and W is the Wronskian of these two solutions:

\beq
W=f_1(t) f_2^\prime (t)-
f_1^\prime (t) f_2(t)= 2 \omega (\omega^2-1) (\omega^2-4)~~.
\label{wr}
\eeq

The usual ambiguity in the solution of a second-order differential
equation, which amounts to the freedom of adding arbitrary linear
combination of the two solutions of the homogeneous equation, is fixed in
the equation (\ref{solut}) by the requirement that for arbitrary $\omega$
the function $g_p(x)$ should have expansion in ascending integer (in fact
integer even) powers of $x$.

Equation (\ref{solut}) demonstrates that the generating function $g_p(x)$
for the propagators $d_n(p)$ is related to the Green function
$G_\omega(t_1,t_2)$ of Ref.\cite{v} by the Fourier transform in time, which
proves the statement$^{\cite{v}}$ about the equivalence of the two
techniques at least at this level of the perturbation theory in $\lambda$.

{}From the equation one can also readily see the nullification of the
on-mass-shell threshold amplitudes for the scattering amplitudes $2 \to n$.
Indeed, this amplitude is given by the residue of the propagator $D_n(p)$ at
the double pole, when both the final momentum $p$ and the initial $p+nq$
are on the mass shell and $p_0=\epsilon$ is negative. Therefore one can set
$\epsilon$ at its on-mass-shell value, $\epsilon=-n/2$, and look for the
double pole at $\omega=n/2$. (Thus the difference $\omega - n/2$ can be used
as a measure of how far the incoming particles are off the mass shell.) The
integration in eq.(\ref{solut}) however produces only {\it single} poles at
$\omega = \pm (\epsilon +2k)$ with integer $k$, and the missing single pole
can come only from a zero of the Wronskian (\ref{wr}).  Thus the tree-level
amplitude of the on-mass-shell scattering $2 \to n$ is non-vanishing only
for $\omega=1$ and $\omega=2$, i.e. for $n=2$ and $n=4$.

I believe that the presented here calculation based on the recursion
relation (\ref{rec1}) for the propagators $D_n(p)$ somewhat clarifies the
relation between this approach and the technique suggested by Brown and also
can be helpful for an interpretation of the further calculations within
Brown's technique.

As a final remark it can be noticed that the nullification of the
on-mass-shell threshold amplitudes arises due to the special value, namely
6, of the coefficient of $({\rm cosh}\,t)^{-2}$ in the differential operator
in the equation (\ref{difurt}). It is well known that when this coefficient
is equal to $N (N+1)$ with positive integer $N$ the operator has $N$
eigenvalues: $\omega=1,\,2, \ldots,N$ and the solutions of the homogeneous
equation at arbitrary $\omega$ have the form of rational functions of $u(t)$
times $u^{\pm \omega}$. This coefficient is theory-dependent, e.g. in a
similar analysis in the Sine-Gordon theory the coefficient of $({\rm
cosh}\,t)^{-2}$ in the analog of eq.(\ref{difurt}) would be equal to 2, so
that $N=1$.  One can readily write a theory with several bosonic fields
where this coefficient can take arbitrary value depending on the coupling
constants.  From a simple generalization of the present analysis one thus
concludes that every time this coefficient takes the value $N(N+1)$ the
tree-level amplitudes of the scattering $2 \to n$ should vanish at the
threshold for $n > 2N$, though the nullification can also occur at smaller
$n$ for other reasons$^{\cite{smith}}$.

I am thankful to Peter Arnold, whose questions and remarks have
stimulated writing this comment.  This work is supported in part by the DOE
grant DE-AC02-83ER40105.

{\Large \bf Figure captions}\\[0.3in]
{\bf Fig. 1}. The propagator $D_n(p)$ with emission of $n$ on-mass-shell
particles all being at rest. The circle represents the sum of all tree
graphs. \\[0.15in]
{\bf Fig. 2}. The recursion equation (\ref{rec1}) for the propagators
$D_n(p)$. The circles correspond to the sums of all tree graphs.

\newpage
\unitlength=1.00mm
\linethickness{0.8pt}
\begin{picture}(130.00,131.00)
\put(75.00,114.00){\circle{14.00}}
\put(50.00,114.00){\vector(1,0){18.00}}
\put(82.00,114.00){\vector(1,0){18.00}}
\put(75.00,121.00){\line(0,1){6.00}}
\put(78.00,120.00){\line(1,2){3.00}}
\put(72.00,120.00){\line(-1,2){3.00}}
\put(75.00,131.00){\makebox(0,0)[cc]{{\large $n$}}}
\put(90.00,110.00){\makebox(0,0)[cc]{{\large $p$}}}
\put(58.00,110.00){\makebox(0,0)[cc]{{\large $p + nq$}}}
\put(48.00,49.00){\circle{14.00}}
\put(48.00,56.00){\line(0,1){6.00}}
\put(51.00,55.00){\line(1,2){3.00}}
\put(45.00,55.00){\line(-1,2){3.00}}
\put(48.00,66.00){\makebox(0,0)[cc]{{\large $n$}}}
\put(63.00,45.00){\makebox(0,0)[cc]{{\large $p$}}}
\put(113.00,49.00){\circle{14.00}}
\put(113.00,56.00){\line(0,1){6.00}}
\put(116.00,55.00){\line(1,2){3.00}}
\put(110.00,55.00){\line(-1,2){3.00}}
\put(113.00,66.00){\makebox(0,0)[cc]{{\large $n_3$}}}
\put(128.00,45.00){\makebox(0,0)[cc]{{\large $p$}}}
\put(55.00,49.00){\vector(1,0){11.00}}
\put(120.00,49.00){\vector(1,0){10.00}}
\put(85.00,49.00){\line(1,0){21.00}}
\put(95.00,42.00){\line(0,1){14.00}}
\put(95.00,49.00){\circle*{2.00}}
\put(95.00,61.00){\circle{10.00}}
\put(95.00,37.00){\circle{10.00}}
\put(95.00,66.00){\line(0,1){3.00}}
\put(92.00,65.00){\line(-1,4){1.00}}
\put(95.00,32.00){\line(0,-1){3.00}}
\put(98.00,65.00){\line(1,2){2.00}}
\put(98.00,33.00){\line(1,-2){2.00}}
\put(92.00,33.00){\line(-1,-2){2.00}}
\put(28.00,49.00){\vector(1,0){13.00}}
\put(95.00,73.00){\makebox(0,0)[cc]{{\large $n_2$}}}
\put(95.00,24.00){\makebox(0,0)[cc]{{\large $n_1$}}}
\put(77.00,50.00){\makebox(0,0)[cc]{{\LARGE $= \Sigma$}}}
\put(75.00,95.00){\makebox(0,0)[cc]{{\bf Figure 1}}}
\put(75.00,11.00){\makebox(0,0)[cc]{{\bf Figure 2}}}
\end{picture}
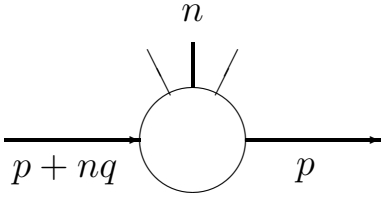
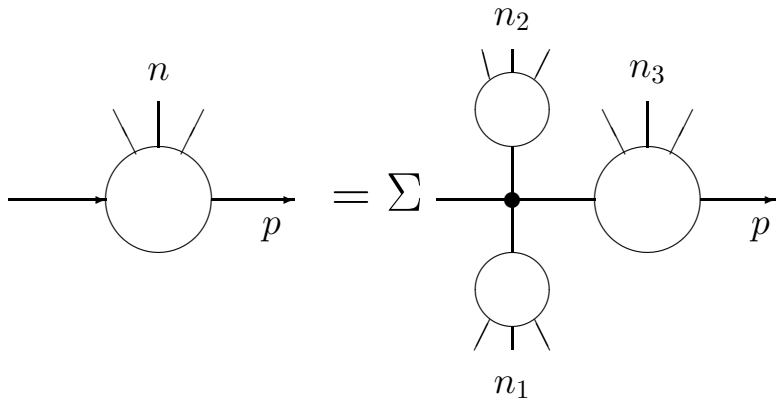
\end{document}